\documentclass[unnumsec,webpdf,contemporary,large]{oup-authoring-template}%

\graphicspath{{Fig/}}

\theoremstyle{thmstyleone}%
\theoremstyle{thmstyletwo}%
\theoremstyle{thmstylethree}%

\begin{document}

\journaltitle{Bioinformatics}
\DOI{DOI HERE}
\copyrightyear{2022}
\pubyear{2019}
\access{Advance Access Publication Date: Day Month Year}
\appnotes{Paper}

\firstpage{1}


\title[Short Article Title]{A large dataset curation and benchmark for drug target interaction}

\author[$\ast$]{Alex Golts}
\author{Vadim Ratner}
\author{Yoel Shoshan}
\author{Moshe Raboh}
\author{Sagi Polaczek}
\author{Michal Ozery-Flato}
\author{Daniel Shats}
\author{Liam Hazan}
\author{Sivan Ravid}
\author{Efrat Hexter}

\authormark{Author Name et al.}

\address{IBM Research - Israel}

\corresp[$\ast$]{Corresponding author. \href{email:alex.golts@ibm.com}{alex.golts@ibm.com}}

\received{Date}{0}{Year}
\revised{Date}{0}{Year}
\accepted{Date}{0}{Year}



\abstract{Bioactivity data plays a key role in drug discovery and repurposing. The resource-demanding nature of \textit{in vitro} and \textit{in vivo} experiments, as well as the recent advances in data-driven computational biochemistry research, highlight the importance of \textit{in silico} drug target interaction (DTI) prediction approaches. While numerous large public bioactivity data sources exist, research in the field could benefit from better standardization of existing data resources. At present, different research works that share similar goals are often difficult to compare properly because of different choices of data sources and train/validation/test split strategies. Additionally, many works are based on small data subsets, leading to results and insights of possible limited validity. In this paper we propose a way to standardize and represent efficiently a very large dataset curated from multiple public sources, split the data into train, validation and test sets based on different meaningful strategies, and provide a concrete evaluation protocol to accomplish a benchmark. We analyze the proposed data curation, prove its usefulness and validate the proposed benchmark through experimental studies based on an existing neural network model.}


\maketitle

\section{Introduction}
Information about small molecules and their interaction with protein targets is invaluable for drug discovery and design, as well as drug repurposing processes. \cite{isigkeit2022consensus, paul2021artificial}
Most drugs demonstrate efficacy via \textit{in-vivo} interactions with their target molecules such as enzymes, ion channels, nuclear receptors and G protein-coupled receptors (GPCRs). However, the experimental validation of drug-target pairs involves costly, time-consuming and challenging work.
Specifically, the discovery process of each New Molecular Entity (NME) costs roughly 1.8 billion dollars\cite{paul2010improve} and the approval of a New Drug Application (NDA) takes between 9 and 12 years \cite{dickson2004key}.
This has motivated researchers to develop data-driven DTI prediction methods as they enable the screening of new drug candidates effectively and efficiently.

Using computational methods, researchers can screen large numbers of compounds and prioritize those that are most likely to be effective and safe. By identifying and optimizing drug candidates using data-driven approaches, researchers can accelerate the drug discovery process and improve the chances of success in clinical trials.  
Identifying new targets for existing or abandoned drugs, namely drug repositioning or repurposing, is another important part of drug discovery.

In this work, we propose a new standard for representing DTI data that can originate from different sources, preparing it for learning-based computational methods via meaningful split strategies, as well as defining the evaluation process meant to compare between competing methods. The ensemble of these three components is typically referred to as a ``benchmark". 
Our proposed DTI benchmark is fully reproducible. We release as open source the code for running the evaluation process, as well as training a model that we use in this paper to demonstrate the benchmark \cite{fusedrug_plmdti}. We also release the data curated for this benchmark as a stand alone artifact~\cite{golts_alex_2023_8105617}.

Information on small molecules and their biological activity can be extracted from scientific publications. Large generic databases such as ChEMBL \cite{zdrazil2023chembl} and PubChem \cite{kim2023pubchem} curate such reported data. More specialized repositories like BindingDB \cite{gilson2016bindingdb} focus on compounds that were already determined to be of ``high quality". Since different such sources have different objectives, there are differences in the data contained within them, the means by which they were collected, the level of details etc'. 
Data-driven Artificial Intelligence (AI) and Machine Learning (ML) approaches are promising in many aspects of drug discovery and development \cite{paul2021artificial}. Therefore it is valuable to create a combined dataset containing all publicly available information on compounds and their bioactivity. Such a large combined dataset should represent the data in a standardized and efficient way, as well as account for overlaps between different source datasets. To this end, we propose a simple but efficient form of representation (Sec.~\ref{sec:representation}). 

Previous efforts were made into combining DTI data from public sources and defining benchmarks. In \cite{isigkeit2022consensus} data from five public sources was used and standardized in a proposed unified representation. Possibly due to the choice of how to access each data source, the resulting number of activities in this curation is only around 20\% compared to our effort. Also of note is that while this work focuses on analysis and insights into the data and its properties, it lacks demonstration of applicability to a concrete task via experiments.  
Even more common are data curations that are at least two orders of magnitude smaller in size compared to ours. MoleculeNet \cite{wu2018moleculenet} is a benchmark specialized in testing machine learning methods related to molecular properties. MUV \cite{rohrer2009maximum} is an example of a bioactivity dataset within it, that is popular in the literature. The goal of our curation is to be as large as possible for the purpose of leveraging the data driven nature of large machine learning models. A dataset like MUV is meant to be small and focused, for the purpose of efficient virtual screening and speeding up biological testing. Accordingly, it contains around 250k DTI pairs. 
Therapeautics Data Commons \cite{huang2021therapeutics} is a collection of datasets and tasks, forming benchmarks in the drug discovery and development space. For DTI, it offers benchmarks based on three datasets: BindingDB that is also a part of our curation, DAVIS and KIBA. Collectively, these benchmarks contain around 200k samples.

In the next sections we elaborate on the public data sources that we used and curation process, define our proposed standardized data representation, analyze the curated data statistics, elaborate on the train/validation/test split logic that we implemented and define evaluation metrics for our proposed benchmark. We then report experiments done with the curated dataset and a baseline DTI model and discuss implementation details as well as enhancements and analyses to the potential and utility of the dataset. 

\section{Data}\label{sec:data}
Next, we describe the sources from which we curate the data, the concrete method of curation, and finally we define a format of standardized representation which we generated and published
\subsection{Sources}\label{sec:sources}
\subsubsection{PubChem}\label{sec:pubchem}
PubChem \cite{kim2023pubchem} is a very large collection that includes information about over 100M compounds and 300M bioactivities. It contains information about chemical and physical properties, biological activities, information about safety and toxicity, patents, literature citations and more.
PubChem offers multiple means for accessing the information including a web interface, programmatic APIs and an FTP server.
For our curation we found it convenient to use the large \textit{bioactivities.tsv} and \textit{bioassays.tsv} files found on the FTP server. For some information that was missing in those files such as the units of measurement for the bioactivity values, we used the corresponding RDF files.
In order to read data efficiently from the very large CSV files, especially the bioactivities file which is around 16GB, we load them to a local SQLite. 
\subsubsection{BindingDB}\label{sec:bindingdb}
BindingDB \cite{gilson2016bindingdb} is a public, web-accessible database of measured binding affinities, focusing chiefly on the interactions of proteins considered to be candidate drug-targets with ligands that are small, drug-like molecules. It contains 2.7M bioactivities for over 1M compounds and ~10K targets.  
We use the MySQL database available for download on the BindingDB website to pull information from the dataset. For convenience, we convert it internally to PostgreSQL.
\subsubsection{ChEMBL}\label{sec:chembl}
ChEMBL \cite{zdrazil2023chembl} is a manually curated database of bioactive molecules with drug-like properties. It combines chemical, bioactivity and genomic data to be used by researchers on their path to the discovery of effective new drugs. In terms of bioactivity data, it contains over 20M activity samples for 2.4M compounds and 15k targets. These originate from 1.6M assays.  
Here we use directly the PostgreSQL version of the database available for download on the ChEMBL website to pull data from the dataset.

\subsection{Representation}\label{sec:representation}
To represent DTI data from multiple sources in a unified and efficient way we define three Tab Separated Value (TSV) table files, referred to as \textit{pairs}, \textit{ligands} and \textit{targets} files. 
The \textit{pairs} file contains a row for each target-ligand activity sample. To avoid redundancy and facilitate efficiency, the \textit{pairs} file does not contain the raw sequences of the ligand/compound and target protein. Instead it contains IDs for both entities, that will correspond to columns/fields with the same name in the \textit{ligands} and \textit{targets} files, which in turn contain the sequence information for each unique ligand and target, respectively. 

\subsubsection{Pairs file}\label{sec:pairs_file}
In this file, every row is a sample pair, and the columns/fields are as follows:
\begin{enumerate}
  \item \textbf{source\_dataset\_versioned\_name} - identifier that specifies the source dataset and version, e.g. ``bindingdb\_14-04-2022".
  \item \textbf{source\_dataset\_activity\_id} - an identifier for each sample in the source dataset.
  \item \textbf{source\_dataset\_target\_id} - an identifier for each target in the source dataset.
  \item \textbf{ligand\_id} - unique identifier for the ligand. Serves as a foreign key to the Ligands table.
  \item \textbf{target\_id} - UniprotKB \cite{uniprot2019uniprot} accession identifier for the target protein.
  \item \textbf{activity\_type} - Type of the assay/measured activity, e.g. IC50, $K_{d}$ etc'.
  \item \textbf{units} - activity measurement units, e.g. nM.
  \item \textbf{activity\_value} - quantitative value of the activity measurement.
  \item \textbf{activity\_label} - qualitative activity label, e.g. Active, Inactive etc'.
  \item \textbf{relation} - some assays report a relation w.r.t. the activity value, e.g. $<=$, $<$, etc'.
  \item \textbf{source\_dataset\_assay\_id} - identifier for the assay within the source dataset.
  \item \textbf{source\_description} - optional description provided within the source dataset. could be a patent ID, DOI, or any description of the conducted assay.
  \item \textbf{date} - a date that was reported as part of the assay.
  \item any other required custom fields.
\end{enumerate}

\subsubsection{Ligands file}\label{sec:ligands_file}
In this file, each row is a unique compound/ligand in the dataset, and the columns/fields are as follows:
\begin{enumerate}
    \item \textbf{ligand\_id} - unique identifier for a ligand/compound. serves as a primary key for this file, referred to from a foreign key in the pairs file.
    \item \textbf{inchi} - InChI identifier for the compound/ligand.
    \item \textbf{canonical\_smiles} - canonical SMILES representation of the compound/ligand.
    \item \textbf{\{source\_dataset\_versioned\_name\}\_ligand\_id} - identifier for the compound/ligand in the source dataset. The column name will depend on the source dataset, e.g. ``bindingdb\_14-04-2022\_ligand\_id".
    \item any other required custom fields.
\end{enumerate}

Lending terminology from relational databases, the \textit{ligand\_id} field can be thought of as the primary key in the \textit{ligands} file/table. It identifies uniquely every ligand/compound in the combined dataset. In the \textit{pairs} file/table (Sec.~\ref{sec:pairs_file}), a field with the same name serves as a foreign key which refers to the primary key in the \textit{ligands} file/table.

\subsubsection{Targets file}\label{sec:targets_file}
In this file, each row is a unique protein target in the dataset, and the columns/fields are as follows:
\begin{enumerate}
    \item \textbf{target\_id} - UniprotKB accession identifier for the target protein. It corresponds to the column with the same name in the pairs file, although we do not strictly impose it to be a unique identifier (for some targets it is not available, in which case another identifier needs to be used).
    \item \textbf{canonical\_aa\_sequence} - amino acid sequence representing the protein target.
    \item \textbf{\{source\_dataset\_versioned\_name\}\_target\_id} - identifier for the target in the source dataset. The column name will depend on the source dataset, e.g. ``bindingdb\_14-04-2022\_target\_id".
    \item any other required custom fields.
\end{enumerate}

When constructing a combined dataset from multiple data sources, we clean the data from each individual data source by removing duplicate pairs, and ligands or targets with missing sequence information. We then re-apply the clean up logic to the merged/combined dataset.
Additionally, we create two versions for the combined dataset, a standard one and a \textit{native} one. In the latter, each data source will contain only samples originating from that source.  
In practice, samples in any original data source, may contain samples curated from other data sources. The information about this is retained, so in the \textit{native} version, we filter out such samples. The motivation for this is to reduce the intersection and redundancy between data sources, and operating under the assumption that each data source likely contains the "best" version of the data that originate from within itself, and potentially a sub-optimal/not up-to-date version of data from other sources. 

\subsection{Analysis}\label{sec:representation}
In this section we perform quantitative analysis of the curated benchmark data, to gain some insights about its properties. 

\begin{table*}[t]
\caption{Dataset sample and qualitative labels statistics.}
\tabcolsep=0pt
\begin{tabular*}{\textwidth}{@{\extracolsep{\fill}}lcclcc@{\extracolsep{\fill}}}
\toprule%
  &\multicolumn{2}{@{}c@{}}{Samples distribution} & & \multicolumn{2}{@{}c@{}}{Outcome distribution} \\
\cline{2-3}\cline{5-6}%
Data source & \# & \%  & Outcome & \# & \% \\
\midrule
BindingDB  & 1017160 & 1.17 & Active & 929656 & 1.07 \\
ChEMBL  & 451806 & 0.52  & Inactive & 83703190 & 95.98 \\
PubChem & 85742394 & 98.31 & Inconclusive & 1529908 & 1.75\\
\bf{Total}& \bf{87211360} & \bf{100}&Other & 1048606 & 1.2\\

\botrule
\end{tabular*}
\begin{tablenotes}%
\vspace*{6pt}
\end{tablenotes}
\label{tab:number_of_samples}
\end{table*}
Tab.~\ref{tab:number_of_samples} shows the numbers of samples in the curated dataset, and their distribution across sources, as well as the distribution of qualitative labels.
We see that PubChem accounts for the vast majority of the data in the curation, and that most of the samples represent non binding pairs. 

Tab.~\ref{tab:types_of_activities} shows the distribution of samples with respect to different activity types measured in them. For simplicity, we only include those activity types which account for at least 0.1\% of the data. 

\begin{table}[!t]
\caption{Distribution of the types of activities measured}%
\begin{tabular*}{\columnwidth}{@{\extracolsep\fill}llll@{\extracolsep\fill}}
\toprule
Activity type & \#  & \% \\
\midrule
Potency    & 20694972   & 73.02 \\
IC50    & 5532330   & 19.52 \\
EC50    & 1573837   & 5.55\\
$K_{i}$ & 280461 & 0.99 \\
$K_{d}$ & 93275 & 0.33 \\
Inhibition & 62376 & 0.22 \\
Activity & 57660 & 0.20 \\
\% Control & 47753 & 0.17 \\
\bf{Total} & \bf{28342664} & \bf{100} \\
\botrule
\end{tabular*}
\label{tab:types_of_activities}
\end{table}

\section{Split logic}\label{sec:split logic}
We perform three types of splits: \textit{lenient}, \textit{cold-ligand} and \textit{cold-target}. 

\textit{Lenient} split requires that a data entry (i.e. ligand and target pair) is only present in one split set (either train, validation or test). This is achieved by randomly splitting the set of unique pairs, with probabilities depending on the required set size. In our case we used 70\% of the data for train, 15\% for validation and 15\% for test.

\textit{Cold-ligand} split requires that a given compound/ligand is only present in one split set. This is achieved by randomly splitting the set of unique ligands according to the required train/validation/test proportions. Some ligands may have more corresponding targets than others, which may result in final set sizes that differ from the requirement. We haven't noticed such behavior when using our large curated dataset. However, it may require consideration when using smaller datasets containing fewer ligands.

\textit{Cold-target} split requires that a given target is only present in a single split. This is achieved by randomly splitting the set of unique targets. As with \textit{cold-ligand} split, smaller datasets with unbalanced numbers of pair ligands per target may require adjustment of the splitting strategy in order to yield the required split sizes.

We chose to focus on these three types of splits, that account for model robustness and generalization to new entities. We note however the existence of other split choices in the literature as well. These include:
\begin{itemize}
  \item time-based splits - evaluating the generalization to assays performed further into the future compared to those the model was trained on.
  \item scaffold based - similar in concept to the \textit{cold-ligand} split, but making it more restrictive by ensuring ligands with even a similar scaffold/core structure are not present in both train and validation/test sets.
  \item clustering based - similar in concept to \textit{cold-"entity"} splits, but making it more restrictive by ensuring entities with even a similar sequence representation are not present in both train and validation/test sets.
\end{itemize}

\subsection{Implementation details}\label{sec:split implement}
A general split is defined by a set of data attributes (Sec.~\ref{sec:representation}). Combined values of these attributes identify an entity according to which the split is performed. In case of a \textit{lenient} split, an entity is defined by the attributes $\{\bf{target\_id}, \bf{ligand\_id}\}$, and all entries sharing the same attribute values belong to the same entity. In case of a \textit{cold-target} split, an entity is defined by a single attribute, \textbf{target\_id}. The total set of entities within the dataset is then split into train, validation and test subsets, such that any sample from a given entity belongs to one subset only. 

This approach allows the definition of more complex entities, as well as split hierarchies. In our benchmark, the dataset is split into train, validation and test sets according to the main split strategy (\textit{lenient}, \textit{cold-ligand} and \textit{cold target}), with the train set further split into 5 \textit{lenient} folds.

\section{Evaluation metric}\label{sec:evaluation metric}
One may utilize the curated DTI dataset in many ways to suit different research purposes. For example, it's possible to use different subsets of the data based on activity types of interest, or any other criteria. It's also possible to suggest different evaluation metrics of interest. 

Here we propose a concrete benchmark implementation. We evaluate it through experiments with a specific model architecture in the next section.

In our proposed benchmark we solve a binary classification task. For that purpose, we use the ~87M samples that have qualitative labels (see Tab.~\ref{tab:number_of_samples}) originating from PubChem (Sec.~\ref{sec:pubchem}). We filter out samples with ``Inconclusive" labels and only keep those labeled ``Active" or ``Inactive". As evaluation metrics we use Area Under Receiver Operating Characteristic (AUROC) and Area Under Precision-Recall (AUPR), both well recognized metrics for classification tasks.
Our data is highly imbalanced with the ``Inactive" class more prevalent by a significant margin. Our split strategies keep this imbalance in the test set as well, to reflect more accurately the actual data distribution. Therefore, one should note that AUPR performance is expected to be lower than AUROC \cite{davis2006relationship}. This is because while in AUROC the ``baseline" for a binary classifier that reflects ``random" performance is 50\%, the expected AUPR for such classifier is the positive class prevalence. 
Tab.~\ref{tab:positive_class_prevalence} shows the prevalence of the positive ("Active") class in the test set for the three split strategies that we evaluated.
\label{tab:positive_class_prevalence}
\begin{table}[!t]
\caption{Positive class prevalence}%
\begin{tabular*}{\columnwidth}{@{\extracolsep\fill}llll@{\extracolsep\fill}}
\toprule
Split strategy & \# of samples  & \# of positives & positive margin [\%] \\
\midrule
Lenient    & 12644133  & 139476  & 1.103 \\
Cold-ligand    & 12318559  & 156350  & 1.269 \\
Cold-target    & 12318548 & 156350 & 1.269 \\
\botrule
\end{tabular*}
\end{table}

\section{Experiments}\label{sec:experiments}
As a baseline DTI model to apply to our curated dataset we train a model with an architecture as the one published in \cite{sledzieski2022adapting}. The model, referred to as PLM-DTI, leverages a large pre trained Protein Language Model (PLM) to obtain state of the art results on multiple small DTI datasets. We open source our implementation \cite{fusedrug_plmdti} which is based on the original implementation of \cite{sledzieski2022adapting} found in ~\cite{conplex_dev}, adding code for handling our curated dataset, performance evaluation, as well as some enhancements as discussed later on.

\subsection{Model}\label{sec:model}
The PLM-DTI model uses a pretrained large PLM to encode the protein target sequence, and Morgan Fingerprints \cite{morgan1965generation} to encode the ligand SMILES representation.
It then transforms the ligand and target encoding into a shared latent space using separate fully connected layers. Then, given two latent embeddings, the probability of an interaction is computed as the cosine similarity between the ligand and target embedding vectors.

\subsection{Training details}\label{sec:training details}
The model was trained using Adam optimizer \cite{kingma2014adam} with a learning rate of $10^-4$ and batch size of 32, minimizing the focal loss \cite{lin2017focal} between the ground truth and predicted DTI probabilities. We used different train/validation/test split strategies (Sec.~\ref{sec:split logic}), all of which devoted 70\% of the available data for training, 15\% for validation and 15\% for testing. We limited the training time in most of our experiments on our large DTI dataset to 24 hours using a single Nvidia V-100 GPU. This resulted in around 250k iterations, meaning the model saw only roughly 8M samples out of the 61M available for training. It is reasonable to assume that performance could be further improved with more training.

\subsection{Benchmark results}\label{sec:benchmark results}
Tab.~\ref{tab:results} shows AUROC and AUPR curves results for different types of split strategies (Sec~\ref{sec:split logic}). 
\label{tab:results}
\begin{table}[!t]
\caption{Benchmark results}%
\begin{tabular*}{\columnwidth}{@{\extracolsep\fill}llll@{\extracolsep\fill}}
\toprule
Split strategy & AUROC  & AUPR \\
\midrule
Lenient    & 0.843 & 0.145 \\
Cold-ligand    & 0.833 & 0.136 \\
Cold-target    & 0.671 & 0.029 \\
\botrule
\end{tabular*}
\end{table}

\subsection{Focal loss}\label{sec:focal loss}
\begin{figure}
    \centering
    \includegraphics[width=\columnwidth]{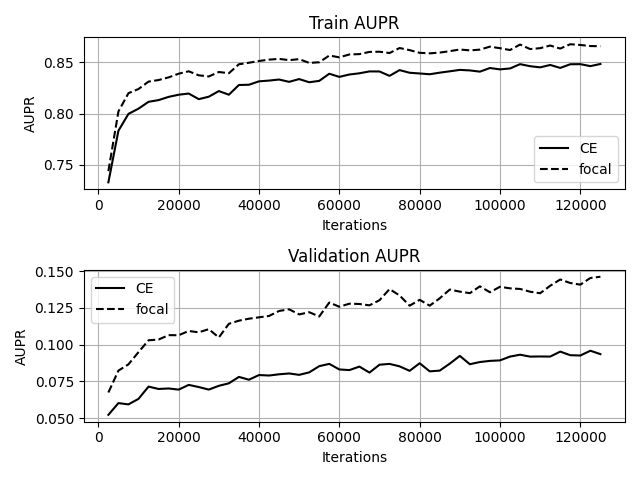}
    \caption{Train and validation AUPR using cross entropy (CE) loss vs. focal loss}
    \label{fig:focal loss trainval performance}
\end{figure}

Given the highly imbalanced nature of the data (Sec.~\ref{sec:evaluation metric}) we opt to use Focal Loss instead of the binary cross entropy loss in the original work. 
In the "lenient" split benchmark, using focal loss led to 53\% improvement in the AUPR. The improvement is also evident from Fig.~\ref{fig:focal loss trainval performance} showing the train and validation AUPR as a function of training iterations.
\subsection{Deeper model}\label{sec:deeper model}
The PLM-DTI model (Sec.~\ref{sec:model}) was shown to work well utilizing a pre-trained PLM and a single learned fully connected layer, when trained on a small dataset of interest. In this work we propose a very large dataset curation. We hypothesize that using such large dataset, we could benefit from training more layers. 
We propose a modified PLM-DTI model with three learned fully connected layers instead of one and compare it to the original model using our benchmark. Tab.~\ref{tab:results with deeper model} shows the results obtained with the deeper model. We see a clear improvement in terms of AUPR, most significant in the lenient split. We note that no performance improvement was obtained from using a deeper model on the original small datasets used by \cite{sledzieski2022adapting}. Instead, it led to decreased performance likely due to overfit. This further suggests the potential for true improved DTI performance from using larger datasets.
The improvement obtained on our large curated dataset demonstrates the potential for advancing drug discovery efforts through utilization of more curated data than commonly used in existing works. This is enabled by adapting existing models, and working to develop new ones more suited for large datasets.

\begin{table}[!t]
\caption{Benchmark results with deeper model.}%
\begin{tabular*}{\columnwidth}{@{\extracolsep\fill}lllll@{\extracolsep\fill}}
\toprule
&\multicolumn{2}{@{}c@{}}{AUROC} & \multicolumn{2}{@{}c@{}}{AUPR}\\
\cline{2-3}\cline{4-5}%
Split strategy & value  & gain [\%] & value & gain [\%] \\
\midrule
Lenient    & 0.856 & 1.5 & 0.207 & 42.7\\
Cold-ligand    & 0.845 & 1.4 & 0.187 & 37.4\\
Cold-target    & 0.623 & -7.1 & 0.031 & 6.8\\
\botrule
\end{tabular*}
\label{tab:results with deeper model}
\end{table}

\section{Conclusions}\label{sec:conclusions}
We publish a large DTI dataset curated from multiple public sources that can serve as a bedrock for a variety of research efforts in the drug discovery space. We propose an efficient representation for the data, analyzed it and demonstrated its utility across several split types, using a state of the art model that leverages a pretrained PLM. We show the benefit of using focal loss for training on such highly imbalanced datasets, and the potential of such a large dataset in improving DTI performance, provided that deeper models are used. In addition to publishing the data curation for the community, we open source our model training and evaluation code. 

\section{Competing interests}
No competing interest is declared.

\section{Acknowledgments}
The authors thank Joseph A. Morrone from IBM Research for the valuable discussion and suggestions.

\bibliographystyle{abbrv}
\bibliography{references}


\end{document}